\documentclass[11pt,a4paper]{article}
\usepackage[utf8]{inputenc}
\usepackage[english]{babel}
\usepackage{amsmath}
\usepackage{amsfonts}
\usepackage{amssymb}
\usepackage{graphicx}
\usepackage[left=1.54cm,right=1.54cm,top=2.54cm,bottom=2.54cm]{geometry}
\usepackage[numbers]{natbib}
\usepackage{authblk} 

\usepackage{bm}
\usepackage{graphicx,subfigure}
\newcommand{\figref}[1]{Fig.~\ref{#1}}
\newcommand{\avop}[1]{\ensuremath{\langle{#1}\rangle}}

\usepackage{amssymb}
\usepackage{amsthm}
\usepackage{amsfonts}
\usepackage{amsmath}
\usepackage{t1enc}
\usepackage{textcomp}

\usepackage{color}
\usepackage[normalem]{ulem} 

\newcommand{\Replace}[2]{\bgroup\noindent\textcolor{red}{\xout{#1} #2}\egroup\ignorespacesafterend}
\newcommand{\Delete} [1]{\bgroup\noindent\textcolor{red}{\xout{#1}}\egroup\ignorespacesafterend}
\newcommand{\Insert} [1]{\bgroup\noindent\textcolor{red}{#1}\egroup\ignorespacesafterend}
\newcommand{\Comment}[1]{\definecolor{Mygray}{gray}{0.50}\bgroup\color{Mygray}\noindent#1\egroup\ignorespacesafterend}

\newcommand \Michael[1] {\bgroup\noindent[\textcolor{blue}{\textbf{Michael}: #1}]\egroup\ignorespacesafterend}
\newcommand \Stefan [1] {\bgroup\noindent[\textcolor{blue}{\textbf{Stefan}: #1}]\egroup\ignorespacesafterend}
\newcommand \Jane [1] {\bgroup\noindent[\textcolor{blue}{\textbf{Jane}: #1}]\egroup\ignorespacesafterend}
\newcommand \Tom  [1] {\bgroup\noindent[\textcolor{blue}{\textbf{Tom}: #1}]\egroup\ignorespacesafterend}

\title{Propagating compaction bands in confined compression of snow: Experiment and Modelling}

\author[1]{T. W. Barraclough}
\author[1]{J. R. Blackford} 
\author[1,2]{S. Liebenstein}
\author[2]{S. Sandfeld}
\author[1]{T. J. Stratford}
\author[2]{G. Weinl\"ander}
\author[1,2,*]{M. Zaiser}

\affil[1]{School of Engineering, The University of Edinburgh, Edinburgh, UK.}
\affil[2]{Department of Materials Science, Institute for Materials Simulation WW8, FAU Erlangen-Nuremberg, Germany.}
\affil[*]{Corresponding author, michael.zaiser@ww.uni-erlangen.de}

\usepackage[doublespacing]{setspace}

\begin{document}

\maketitle

\begin{abstract}
We show that the plastic deformation of snow under uniaxial compression is characterized by complex spatio-temporal strain localization phenomena.  Deformation is characterized by repeated nucleation and propagation of compaction bands. Compaction bands are also observed during the very first stage of compression of solid foams where a single band moves across the sample at approximately constant stress. However, snow differs from these materials as repeated nucleation and propagation of bands occurs throughout the subsequent hardening stage until the end of the deformation experiment. Band nucleation and/or reflection of bands at the sample boundaries are accompanied by stress drops which punctuate the stress strain curve. A constitutive model is proposed which quantitatively reproduces all features of this oscillatory deformation mode. To this end, a well-established compressive plasticity framework for solid foams is generalized to account for shear softening behavior, time dependence of microstructure (`rapid sintering') and non-locality of damage processes in snow.
\end{abstract}

\section{Introduction}

Irreversible deformation of snow plays an important role in a number of problems ranging from the interaction of snow with winter sports equipment \citep{Wu05} over vehicle traction to snowpack stability and avalanche release \citep{Schweizer99}. The strength of snow is strongly influenced by the presence of bonds connecting ice granules in the snow microstructure. As a consequence, shear deformation reduces strength as bonds are broken, but densification increases strength as higher density leads to an increased number of contacts. The formation of bonds between adjacent ice granules can be envisaged as a thermodynamically driven sintering process \citep{Blackford07} which leads to strengthening of snow over time (ageing) \citep{DeMontmollin82}. The interplay between strain softening and age hardening leads to ductile behavior at low strain rates, when broken bonds have sufficient time to re-form, and to quasi-brittle behavior at high strain rates when this is not possible. This transition can be observed both in pure shear \citep{DeMontmollin82} and in compressive deformation \citep{Kinosita67}.

At intermediate strain rates where ageing and softening processes occur on the same time scale, complex spatio-temporal deformation patterns and oscillation phenomena may occur. Such phenomena are well documented in deformation of metals where they have been analysed under the generic term of {\em strain rate softening} instabilities \cite{Zaiser97a,Zaiser97b,Zaiser99} but have never been studied in snow or similar cohesive-granular materials. The present investigation presents the first experimental and computational investigation of spatio-temporal deformation oscillations in snow as observed in compression experiments which are carried out at strain rates that are borderline between the ductile and quasi-brittle deformation regimes. 

\section{Experimental method}

\begin{figure*}[thb]
\centering
\includegraphics[width=0.98\textwidth]{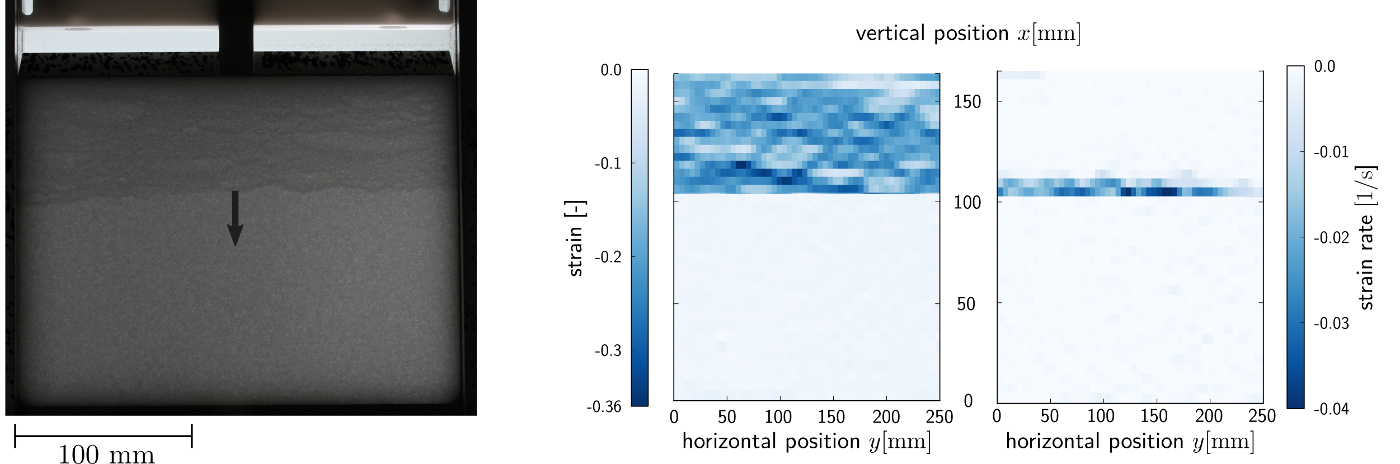}
\caption{Strain and strain rate pattern in a specimen deformed at $\dot{e}_{\rm ext} = 5.77 \times 10^{-3} s^{-1}$ to 9.94 \% overall compressive strain; left: photographic image  showing the compaction front, the arrow indicates the direction of front propagation; center: corresponding strain distribution as obtained by DIC; right: corresponding strain rate distribution; note that the outermost edges of the sample are not covered by the DIC analysis.}
\label{fig:exp_defo} 
\end{figure*}

Specimens of artificially produced dry snow with density $\rho=370\,\rm{kg/m}^{-3}$ and mean grain size $\xi \approx\,0.2$mm were contained within a rectangular transparent container with Aluminium alloy side walls and a glass front and back. The specimens were compacted from above by an anvil moving at fixed rates of $1.125\,\rm{mm\,s}^{-1}$ and $5\,\rm{mm\,s}^{-1}$, thus providing nominal strain rates of $\dot{e}_{\rm ext} = 5.77 \times 10^{-3} s^{-1}$ and $2.56 \times 10^{-2} s^{-1}$. The experiments were carried out at a temperature of $T=-10^\circ$C. 
During deformation, the driving force was recorded by a load cell located above the anvil. A  18 megapixel camera was located in front of the specimens with illumination provided by a flashgun and diffuser located behind the specimens. Images of the transmitted light were recorded at 0.25 s intervals. When the images are viewed, compacted areas become apparent as darker regions (\figref{fig:exp_defo}, left). It was generally found that compaction occurs in a heterogeneous manner by motion of compaction fronts which separate regions of different density (\figref{fig:exp_defo}, see also results section). To quantify this phenomenon in  terms of displacements, each image series was analyzed using digital image correlation (DIC) software.  Local strain and strain rate tensors were then calculated from the displacement fields. Further details of the experimental set-up are given in the Supplementary Material.

\section{Constitutive Model}
\label{sec:const_model}

To model compressive deformation of snow, we start out from a model proposed by \citet{Zaiser13} who generalize the compressible plasticity model formulated by \citet{Deshpande01} for deformation of solid foams to include non-locality and strain localization phenomena. We modify this model to account for the interplay between shear softening of snow associated with the breaking of bonds between ice granules upon shear deformation \citep{McClung79}, and rapid sintering processes which re-establish intergranular links and restore strength over time. Pore pressure effects, which may influence dynamic snow compaction under impact loads \citep{Wu05}, are not relevant at the presently investigated compaction rates and are therefore not taken into account. 

The stress tensor is denoted as $\bm{\sigma}$ with components $\sigma_{ij}$. It fulfils the momentum balance equation which for quasi-static deformation processes as considered here is $\partial_i \sigma_{ij} = 0$ where we sum over repeated indices. The first invariant of the stress tensor is ${\rm Tr}({\bm \sigma)} = \sigma_{ii}$, and the deviatoric stress is ${\bm{\sigma}}_{\rm dev} = {\bm{\sigma}} - (1/3) {\rm Tr}({\bm{\sigma}}) \bf{1}$ with components $\sigma^{\rm dev}_{ij}$. Our model is based on the Deshpande-Fleck yield criterion which has been shown to give a good description of the deformation of solid foams of low to intermediate density \cite{Deshpande01} and may thus be appropriate for snow where typical relative densities are in the range of 0.1-0.4. It is given by
\begin{equation}
f({\bm{\sigma}}) = \sigma_{\rm eq} - S = 0, \quad\;
\sigma_{\rm eq}^2 = \frac{3\sigma^{\rm dev}_{ij}\sigma^{\rm dev}_{ij}}{2(1+\alpha)}  + \frac{\alpha(\sigma_{ii})^2}{1+\alpha}
\label{Eq:yield}
\end{equation}
where $\sigma_{\rm eq}$  is the equivalent stress and the elastic domain is defined by $f < 0$. The plastic strain rate is assumed to derive from an associated flow rule, $\dot{\bm {\epsilon}}^{\rm pl} = \dot{e} (\partial f /\partial \bm{\sigma})$ where the equivalent strain rate $\dot{e}$ is determined by imposing that, under temporally changing constraints, the system must not enter the domain $f > 0$ . The non-dimensional parameter $\alpha$ in Eq. \ref{Eq:yield} characterizes the degree of volumetric change during deformation. For the limit case $\alpha = 0$, the model reduces to the standard Von Mises plasticity thory. 

We consider deformation at low to intermediate strains and use an additive decomposition of the strain tensor $\bm \epsilon = \bm \epsilon^{\rm pl} + \bm \epsilon^{\rm el}$ into elastic (reversible) and plastic (irreversible) contributions. The elastic strain $\bm \epsilon^{\rm el}$ is related to the stress via Hooke's law where we assume isotropic elastic behavior with Young's modulus $E$ and Poisson number $\nu$. 
The elastic modulus $E$ in general depends on density and may increase during densification. To describe this dependency, we assume a power-law density dependence as typically observed for solid foams \citep{Gibson88},
\begin{equation}
E = E_0 (\rho/\rho_0)^{c} \label{E},
\end{equation}
where $\rho$ is the stress-free density of the snow, $\rho_0$ is a reference density which we take to be the density prior to compaction, and $E_0$ is the corresponding Young's modulus. 

To complete our constitutive framework we need to relate the flow stress $S$ to the plastic strain tensor. We assume $S$ to depend both on the  density of the snow and on its structure. Phenomenologically, also the yield stress of many solid foams has a power-law density dependence \citep{Gibson88}: 
\begin{equation}
S = \Sigma (\rho/\rho_0)^{a} \label{Eq:Ss},
\end{equation}
where the pre-factor $\Sigma$ depends on the snow microstructure and may be affected by structural softening under shear deformations.
The evolution of the stress-free density $\rho$ as a function of plastic strain follows from the continuity equation
$d\rho/dt + \rho {\rm Tr}(\dot{\bm{\epsilon}})^{\rm pl} = 0$ which can be integrated to $\rho = \rho_0 \exp[- \varepsilon^{\rm pl}_V]$ where the volumetric plastic strain is given by $\varepsilon^{\rm pl}_V = {\rm Tr}\bm{\epsilon}^{\rm pl}$. Combining with Eq. \ref{Eq:Ss} gives 
\begin{equation}
S = \Sigma \exp[- a \varepsilon^{\rm pl}_V] \label{S2} ,
\end{equation}
which for negative volumetric strain describes densification-induced strengthening. 

Shear deformation of snow is associated with failure of bonds connecting ice granules and thus reduces the load bearing capacity of the snow microstructure. We note in passing the proposal of  \citet{Heierli06} and \citet{Heierli08} that volumetric compaction may also be associated with (transient) structural softening. However, the simple viewpoint that densification implies strengthening while shear implies softening, which we adopt in the following, turns out to be fully adequate for understanding the experiments. We characterize the magnitude of shear deformation in terms of the equivalent shear strain $\varepsilon^{\rm pl}_{\rm S}$ which relates to the second invariant of the deviatoric part of the plastic strain tensor by $(\varepsilon^{\rm pl}_{\rm S})^2 = (2/3) \epsilon^{\rm pl,dev}_{ij}\epsilon^{\rm pl,dev}_{ij}$. The changes in strength which result from the competing processes of shear-induced bond failure and strength recovery due to rapid sintering processes are described by the phenomenological equations 
\begin{equation}
\Sigma = \Sigma_{\rm R} + \Sigma_{\rm S}\quad,\quad 
\partial_t \Sigma_{\rm S} = - \frac{\dot{\varepsilon}^{\rm pl}_{\rm S}}{\varepsilon_{\rm S}} \Sigma_{\rm S} + \frac{1}{\tau} (\Sigma_{\rm S}^0 - \Sigma_{\rm S}).
\label{Eq:Sigma}
\end{equation}
In these equations, $\Sigma_{\rm R}$ is the residual strength associated with intergranular friction which remains even when all bonds between ice granules have failed, $\Sigma_{\rm S}$ is the structural strength contribution of intergranular bonds, $\varepsilon_{\rm S}$ is a characteristic shear strain associated with bond failure, $\Sigma_{\rm S}^0$ is the limit strength reached after prolonged ageing, and $\tau$ is the characteristic time constant for strength recovery by sintering/ageing processes. 

Owing to the possibility of structural softening, deformation may proceed in a localized manner even if the external loading is homogeneous, and spontaneous strain localization is indeed manifest in our experimental data. In such situations, the mathematical formulation of the deformation problem may become ill-posed, leading to mesh dependence of numerical solutions. To mitigate this problem, we adopt the suggestion of \citet{Aifantis84} of adding a second-order gradient of the plastic shear strain to the yield function, a method which has been proven to be thermodynamically consistent by \citet{Gurtin09} (for alternative regularization methods, see \citep{Forest05}). This leads to the constitutive equation
\begin{equation}
S = (\Sigma_{\rm R} + \Sigma_{\rm S})\exp[- a \varepsilon^{\rm pl}_V] + E_0 l^2 \Delta \varepsilon^{\rm pl}_{\rm S}\;. 
\label{Eq:S}
\end{equation}
Mathematically, the non-local term suppresses strain localization on scales below the characteristic length $l$ and defines a finite, mesh independent width of the deformation bands. The Laplacian in the yield condition necessitates a corresponding higher-order boundary condition, which we take to be $\bm n. \bm{\nabla} \varepsilon^{\rm pl}_{\rm S} = 0$, i.e., we require the gradient of the shear strain to be zero in the direction $\bm n$ perpendicular to the specimen surface. This choice of the higher-order boundary condition is found to correctly reproduce the behavior observed when a deformation band reaches the end of the specimen. 

In applying our model to the compression experiments considered here, we assume the parameters $\nu = 0, \alpha = 1/2$. This allows us to reduce the model to three coupled equations for the axial compressive stress $s := \sigma_{xx}$ where $x$ is the coordinate along the compression axis, the axial strain $e := \varepsilon^{\rm pl}_{xx} < 0$, and for the structural strength $\Sigma_{\rm S}$. A derivation of the simplified equations and a demonstration of the correctness of the assumptions $\nu \approx 0, \alpha \approx 1/2$ is given in the Supplementary Material. The axial stress is found to be 
\begin{equation}
s = E_{\rm eff} \left(e_{\rm ext} - \avop{e}\right),\quad{\rm with}\quad
E_{\rm eff}^{-1} = \avop{ E^{-1}(e(x))}
\label{Eq:sigma}
\end{equation}
where $E_{\rm eff}$ is the effective elastic modulus  of the sample. $\avop{\dots}$ denotes the spatial average along the $x$ axis. The  external strain $e_{\rm ext}$, which is negative, is changed at a constant rate, and the concomitant change in the plastic strain $e$ follows from the yield condition,  
\begin{equation}
s \le (\Sigma_{\rm R} + \Sigma_{\rm S})\exp[- a e] + E_0 l^2 \frac{\partial^2 e}{\partial x^2}\label{Eq:e}
\end{equation}
where $\Sigma_{\rm S}$ evolves according to 
\begin{equation}
\partial_t \Sigma_{\rm S} = \frac{\dot{e}}{\varepsilon_{\rm S}} \Sigma_{\rm S} + \frac{1}{\tau} (\Sigma_{\rm S}^0 - \Sigma_{\rm S}).
\label{Eq:Sigma2}
\end{equation}
From these equations, a necessary condition for structural softening and strain localization derives as $a\varepsilon_{\rm S} \le \Sigma_{\rm S}/(\Sigma_{\rm R} + \Sigma_{\rm S})$. This condition (for derivation and further discussion, see supplementary material) is tantamount to the requirement that the strain softening processes due to shear deformation must outweigh the competing process of densification-induced hardening.  

To solve the system of integrodifferential equations for $e$, $s$ and $\Sigma_{\rm S}$, we start from initial conditions $s = e = 0, \Sigma_{\rm S} = \Sigma_{\rm S}^0$. In each time step we change $e_{\rm ext}$ according to the imposed strain rate and evaluate $s$ and $e$ self-consistently from Eqs. \ref{Eq:sigma} and \ref{Eq:e} with the boundary condition $\partial_x e(0) = \partial_x e(L)=0$. By comparing with the previous time step we determine the local plastic strain rates $\dot{e}$, evaluate the changes in $\Sigma_{\rm S}$ from Eq. \ref{Eq:Sigma2}, and repeat until a given end strain is reached. 

\section{Results}

\subsection{Experimental Observations}

\figref{fig:exp_defo} shows a snapshot of the patterns of axial strain $\epsilon_{xx}(x,y)$ and strain rate $\dot{\epsilon}_{xx}(x,y)$
emerging during a compression test. The plots use Eulerian coordinates, i.e., each volume element is shown at its position in the laboratory frame. It is seen that compaction proceeds in a strongly heterogeneous manner, as a region of increased density at the top of the sample is separated from an uncompacted region at the bottom by a sharp moving front where the strain rate is concentrated. This front is perpendicular to the compression direction. Owing to the intrinsic structural heterogeneity of the snow microstructure, the deformation state behind the front is not completely homogeneous. To facilitate comparison with our model, which provides an effective one-dimensional description, in the following we average the strain and strain rate fields over the $y$ direction. 

\begin{figure*}[htp]
\centering
\includegraphics[width=0.9\textwidth]{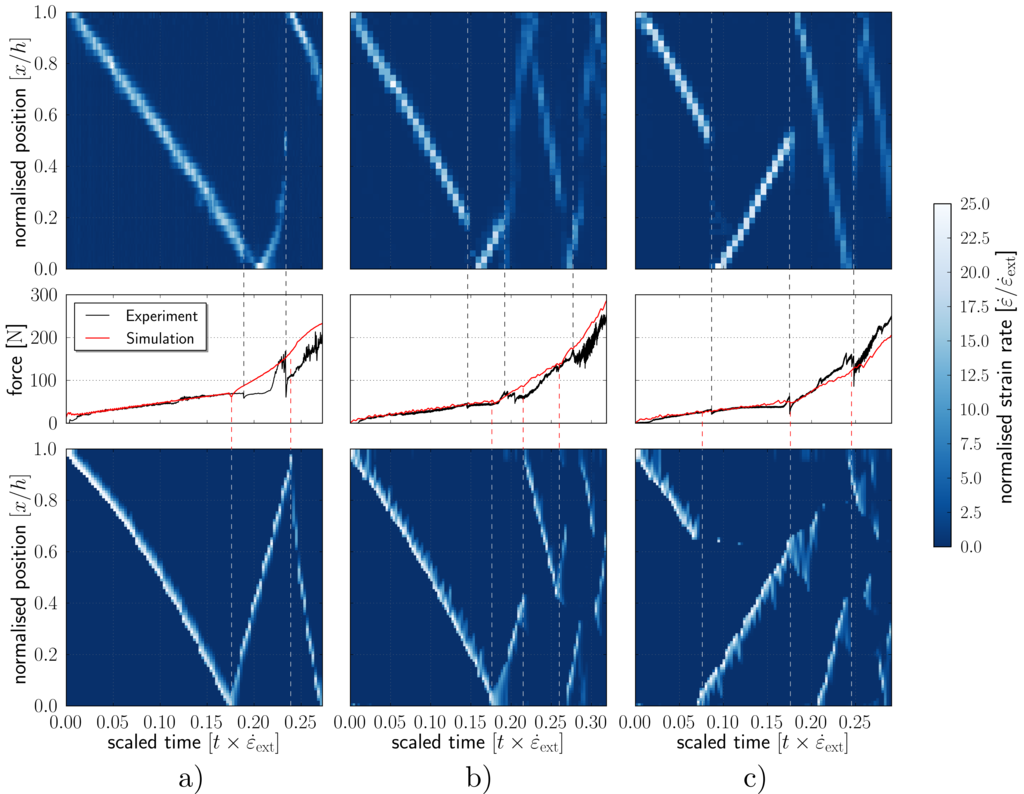}
\caption{Top: space-time plots of the evolution of deformation activity as deduced from DIC; bottom: space-time plots of deformation activity as obtained from simulation; center: corresponding force vs. time/strain curves (black: experiment, red: simulation), the simulated curves have been corrected for friction effects as detailed in the Supplementary Material. (a) represents a sample deformed at $\dot{e}_{\rm ext} = 5.77 \times 10^{-3} $s$^{-1}$, (b,c) samples deformed at $\dot{e}_{\rm ext} = 2.56 \times 10^{-2} $s$^{-1}$; experimental data for deformation of samples (a) and (b) are also shown in Supplementary Movies 1 and 2.}
\label{fig:exp_sim}
\end{figure*}

To visualize band motion, we use space-time plots where we plot the $y$-averaged strain rate $\dot{e}_{\rm tot}(x,t) = \langle \dot{\epsilon}_{xx}(x,y,t) \rangle_y$ in colorscale as a function of $x$ and $t$ (\figref{fig:exp_sim}). In these plots we use Lagrangian $x$ coordinates. i.e., $x$ denotes the position along the compression axis {\em in the initial configuration} (thus the $x$ interval occupied by the sample in the graphs does not shorten during compression). The colour contrast demonstrates the strongly localized nature of deformation. A moving deformation front appears on the space-time plot as an inclined zone of high strain rate. The inclination slope defines the front velocity in the Lagrangian frame. The first front nucleates at the top of the sample and moves at constant (Lagrangian) speed downwards. Once it reaches the bottom of the sample, a new  front nucleates there and moves upwards across the sample until it reaches the top. Repetition of this process leads to a bouncing motion of the locus of deformation. Oscillatory features are also manifest on the force vs. time curves as band nucleation is associated with an up-down oscillation of the driving force (see the dashed lines in \figref{fig:exp_sim} which illustrate the correspondence). Occasionally the bouncing pattern is interrupted as seen in \figref{fig:exp_sim}(c) where the first band gets stuck in the middle of the sample and deformation is taken over by a second band nucleating at the bottom and moving upwards until it merges with the first, whereafter a new band nucleates at the top and the bouncing pattern is resumed. The observations are similar for both investigated deformation rates.

\subsection{Simulation results}

In our simulations we use the following model parameters (for discussion, see the Supplementary Material): Initial elastic modulus $E_0 = 10$ MPa, volumetric change parameter $\alpha = 0.499$, friction coefficient at container surface $\mu =  0.5$, mean initial (homogeneous) strength of snow $\Sigma_0 = 8 \times 10^4$ Pa, exponent of density dependence of elastic modulus $c=3$, exponent of density dependence of strength $a=8$, softening strain $\varepsilon_{\rm S} = 0.05$, residual strength $\Sigma_{\rm R} = 0.05 \Sigma_0$, characteristic time for rapid sintering $\tau = 10$s. The internal length is taken to be $\xi = 0.2$mm, which we also use as spacing of our computational grid. To mimic microstructural heterogeneity, the initial strength at the different grid points is assumed to be Weibull distributed with mean $\Sigma_0$ and Weibull exponent $\beta = 5$. Finally, to mimick the effect of near-surface heterogeneities and stress concentrations due to uneven loading at the specimen ends, structural strength is assumed to decrease towards zero across two surface layers of width $1$mm (5 grid points) and 0.4mm (2 grid points), located at the top and bottom of the specimen, respectively. 

Simulations were carried out at imposed strain rates $\dot{e}_{\rm ext} = 5.7 \times 10^{-3} s^{-1}$ and $\dot{e}_{\rm ext} = 2.5 \times 10^{-2} s^{-1}$, matching the experimental data shown in \figref{fig:exp_sim}. The two runs shown for the higher strain rate differ only with respect to the initial random distribution of local strength.

It is seen that, similar to the experiment, the nucleation of new bands is associated with oscillations in the force vs. time/strain curves, though these are less pronounced than in the experimental counterparts. In the absence of strength fluctuations we consistently observe a bouncing band pattern similar to \figref{fig:exp_sim}(a). Strength fluctuations occasionally induce band arrest (first dashed line in \figref{fig:exp_sim}(b)) and/or intermittent band propagation, as also observed experimentally. Neither in the simulations nor in the experiments do we see a strong influence of the different band propagation modes on the overall shape of the force-displacement curves. 

\begin{figure*}[htb]
\centering
\includegraphics[width=0.9\textwidth]{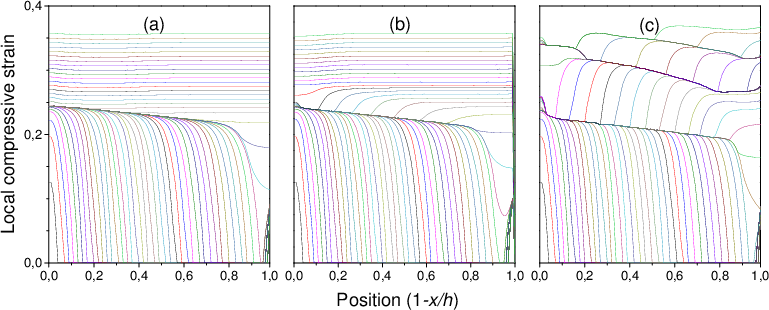}
\caption{Strain pattern evolution for different imposed strain rates, compressive strain 
profiles are plotted as a function of the position along the compression axis
(the top of the sample is at the left of the graphs); adjacent 
profiles differ by equal total strain (time) increments.
(a) $\dot{e}_{\rm ext}\tau = 10$, (b) $\dot{e}_{\rm ext}\tau = 1$, (c) 
$\dot{e}_{\rm ext}\tau = 0.1$, for other parameters see text.}
\label{fig:sim_eps}
\end{figure*}

Varying the strain rate by two orders of magnitude, we see a transition between qualitatively different types of spatio-temporal deformation patterns (\figref{fig:sim_eps}) depending on the value of the product $\dot{e}_{\rm ext}\tau$. For $\dot{e}_{\rm ext}\tau \gg 1$, we find a single compaction band followed by homogeneous deformation, a behavior that is typical for solid foams \citep{Zaiser13}. For $\dot{e}_{\rm ext}\tau = 1$ we see first one secondary band and then, at even lower strain rates, a procession of bands similar to those observed in our experiments (see also the discussion of stability criteria in the Supplementary Material).

\section{Discussion and Conclusions}

We have for the first time demonstrated that the irreversible deformation of snow may proceed in an intrinsically inhomogeneous manner characterized by the repeated nucleation and propagation of localized deformation bands. We could relate this phenomenon to the competition between structural softening under deformation, and age hardening due to rapid sintering processes which restore strength over time. Our findings clearly demonstrate the necessity to investigate and account for strain localization phenomena even when considering initially homogeneous snow samples under homogeneous loads. If shear softening outweights strengthening due to densification, unstable deformation accompanied by spatio-temporal strain localization is bound to occur, otherwise, deformation is expected to proceed in a homogeneous manner associated with monotonic hardening. The difference is manifest not only in laboratory experiments, but in everyday experience: Walking on snow is, in the second case, associated with a foot resistance that increases with increasing penetration depth -- as a consequence, we can {\em walk on} the snow. In the strain softening case, on the other hand, once the critical load needed to nucleate a compaction band is exceeded the foot travels downwards together with the compaction band at constant or even decreasing load -- we have the feeling of {\em breaking through} the snow. This behavior, commonly interpreted in terms of snow heterogeneity ("there is a crust on the snow"), in fact reflects the intrinsic interplay of strain softening and strain localization characteristic even of homogeneous snow. 

The formation of compaction bands has close analogies in deformation of metallic foams, where the initial stage of compressive deformation is often characterised by localized deformation bands which can be explained in terms of localized structural softening (buckling or failure of  cell walls) \citep{Forest05,Zaiser13}. However, repeated nucleation and propagation of deformation bands, which is a conspicuous feature of the present experiments, has never been observed in solid foams. The difference is readily understood if we consider the extremely high homologous temperatures in the present experiments where rapid sintering may lead to strength recovery on short time scales  \citep{Szabo07, Birkeland06}. As a consequence, if the characteristic time required for propagation of a deformation band is on the same order of magnitude as the characteristic time of strength recovery in a snow sample, repeated band nucleation may be observed. 

In the context of snow, the interplay between softening and ageing has been modelled by a number of authors (e.g. \citet{Louchet01,Reiweger09}. However, models published to date cannot describe spatio-temporal strain localization, either because they implicitly assume homogeneous deformation \citep{Louchet01} or because they consider spatial couplings within a mean-field framework which does not account for spatial structure of the deformation field \citep{Reiweger09}. We have demonstrated in the present paper that generalized continuum models (here: gradient plasticity), combined with a simple phenomenological description of strain softening and age hardening processes, offer a framework capable of quantitatively describing the observed spatio-temporal phenomena. Beyond the scope of the present study, such models may offer a generic framework to describe snow deformation also under heterogeneous and time-dependent loading conditions.

\section*{Author contributions}
T.B., T.S. and J.B. designed the apparatus and DIC imaging system, T.B. carried out the experiments, S.S. and S.L. analyzed the data, G.W. wrote the code and performed the simulations, M.Z. formulated and parameterized the model and wrote the manuscript.

\section*{Additional information}
Supplementary information is available in the online version of the paper. 

\bibliographystyle{plain}

\end{document}


\maketitle

\section{Experimental method}

\begin{figure}[thb]
\includegraphics[width=25pc,angle=0,origin=c]{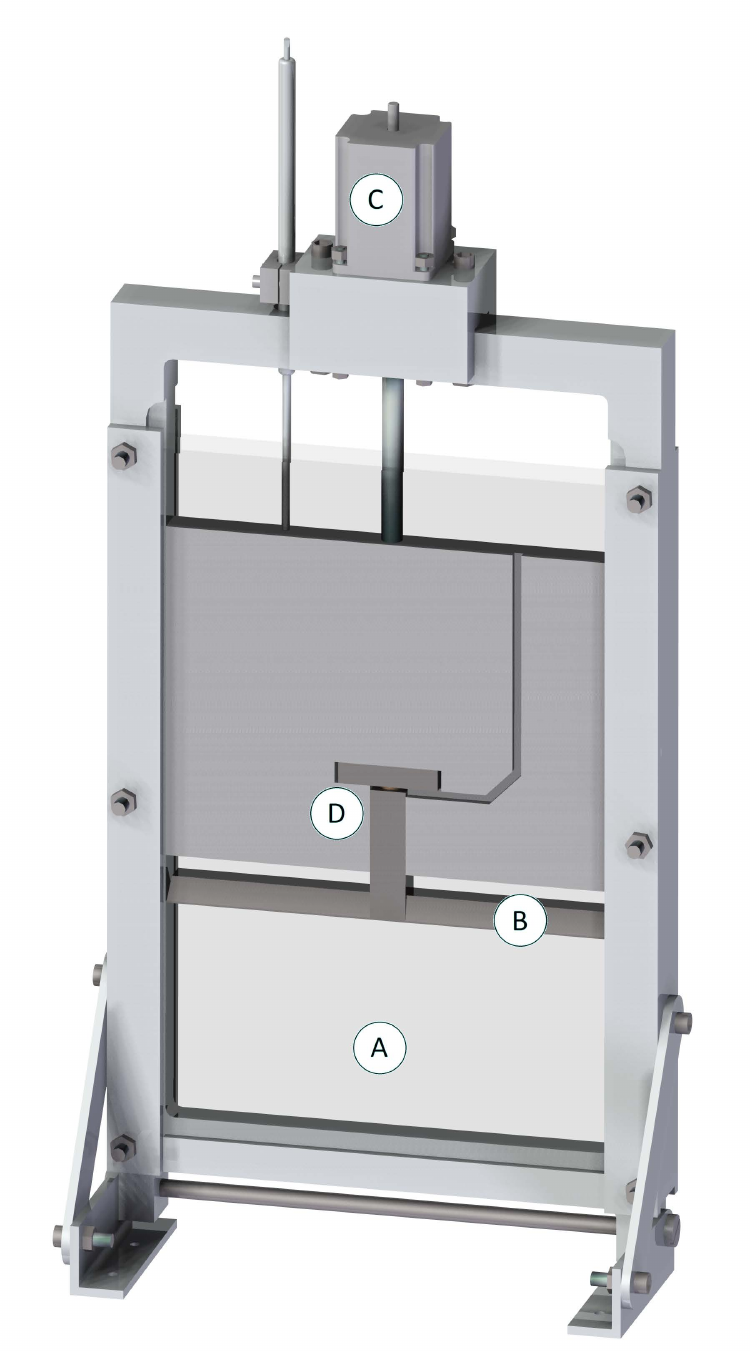}
\caption{Experimental apparatus (schematically). Snow is enclosed in the lower part (a) and compacted by an anvil (b) driven by motor (c). Force is measured by a load cell (d); lateral forces can be determined by measuring the bending of the front/back plate in (a).}
\label{fig:figure1}
\end{figure}

Ice chips were cooled to $-80^\circ$C and crushed into a dry powdery material using a standardised process \cite{Blackford14}. This material was then placed in a covered container in the cold lab for four hours to allow it to reach the temperature of $-10^\circ$C at which the deformation experiments were carried out. This resulted in a cohesive aggregate of ice granules with density $\rho=370\,\rm{kg/m}^{-3}$ and grain sizes ranging from $0.1$ to about $0.5$\,mm (mean grain size $\xi \approx 0.2$\,mm). As is typical for laboratory- or machine-made snow, the density of our samples is at the upper end of the range observed for natural snow. The artificial snow thus produced was sifted through a 1.68\,mm aperture sieve into a flat rectangular box with aluminium alloy side walls and a glass bottom.
Excess snow was removed with a straight edge and the box was closed with a second sheet of glass screwed on top of the specimen.  The container was then rotated upright to provide a specimen $w = 250$\,mm wide, $L = 195$\,mm high and $d = 20$\,mm thick, enclosed in a container with transparent front and back walls within an apparatus as shown in  \figref{fig:figure1}. The total time from sieving to deformation was approximately 5 minutes.

The specimen was compacted by an anvil moving downwards at fixed rate. During deformation, the driving force was recorded by a load cell located above the anvil. The recorded force data were filtered to remove high frequency oscillations associated with the motor drive. Typical compressive forces at the end of a deformation experiment (compressive strain $\varepsilon_{xx} \approx 0.3$) were of the order of 250\,N. 
As the specimen is compacted, it exerts lateral forces on the side walls of the container. These forces were evaluated by determining, at the end of a deformation experiment, the sideways deflection of the container side walls using a pair of dial test indicators, 
and comparing with calculations made under the assumption that the response of the side walls is that of an  elastic thin plate, with elastic properties determined by a calibration experiment. This yielded a total lateral force of about 45\,N at compressive strain $\varepsilon_{xx} = 0.3$. 

An 18 megapixel camera was located in front of the specimen with illumination provided by a flashgun and diffuser located behind the specimen. Images of the transmitted light were recorded at 0.25\,s intervals. To analyze spatio-temporal deformation patterns, each image series was analysed using the digital image correlation software GeoPIV \cite{White03}. This software takes samples of an image (known as patches) and searches for them in subsequent images, thus producing a displacement vector for each patch. The software is capable of locating a patch with sub-pixel accuracy. When this software was used to track small translations of an undeformed specimen, the mean error was found to be significantly less than $40\,\mu$m. The images were analyzed sequentially, using a grid of 64x64 pixel patches spaced at  4.5\,mm and with the patch positions reset to this grid for each image.
Local strain tensors were then calculated from the displacement fields by taking the tensorial displacement gradient which was evaluated using the ParaView package \cite{Henderson07}, and strain rate tensors were evaluated from strain tensors pertaining to sequential images in terms of finite differences.

\section{Application of the constitutive model to the compression experiments}

The general constitutive framework has been defined in the main paper. When applying this framework to the compression experiments,  we make the simplifying assumptions that $\alpha = 1/2$ and $\nu = 0$. These assumptions are tantamout to assuming that, under uniaxial compression, the material exhibits neither elastic nor plastic expansion in the lateral direction. In other words, in this approximation the snow does not interact with the container side walls and the confined snow behaves exactly as a free standing column where we are dealing with a purely uniaxial stress {\em and} strain state. The accuracy of these assumptions will be demonstrated in the section on model parameters below.  

With $\alpha = 1/2$ the equivalent stress is given by
\begin{equation}
\sigma_{\rm eq}^2 = \sigma^{\rm dev}_{ij}\sigma^{\rm dev}_{ij}  + \frac{(\sigma_{ii})^2}{3}
\label{Eq:yield}
\end{equation}
The plastic strain rate as derived from the associated flow rule is   
\begin{equation} 
\dot{\bm {\epsilon}}^{\rm pl} =  \dot{e} \frac{1}{\sigma_{\rm eq}}\left( {\bm \sigma}_{\rm dev} + \frac{1}{3} {\rm Tr}({\bm\sigma}) {\bf 1} \right). 
\label{Eq:flow}
\end{equation}
Identifying the compression direction with the $x$ axis direction of a Cartesian coordinate system and using $\nu \approx 0$, the stress, elastic and plastic strain tensors then have the simple structure 
\begin{equation}
{\bm{\sigma}}
= s \left(
\begin{array}{lll} 1 & 0 & 0 \\
                           0 & 0 & 0 \\
                           0 & 0 & 0 \\
\end{array}\right)
= E(e(x)) {\bm \varepsilon}^{\rm el}(x) \,,\quad
{\bm \varepsilon}^{\rm pl} = e(x) \left( \begin{array}{lll} 1 & 0 & 0 \\
                           0 & 0 & 0 \\
                          0 & 0 & 0 \\
\end{array}\right).\label{se}
\end{equation}
Owing to the parameter choices $\alpha = 1/2$ and $\nu = 0$ there is no lateral expansion or contraction of the sample and hence no interaction with the confining walls. The measured data for the axial compaction forces and lateral forces allow us to assess the quality of this approximation: At the end of each deformation experiment we find typical axial forces of 250\,N which, given the sample geometry, translates into a typical axial stress $\sigma_{xx} = 5 \times 10^4$ Pa. With an end height of the sample of $L' \approx 140$\,mm, the typical lateral force of 45\,N translates into a typical lateral stress of $\sigma_{zz} = 12 \times 10^2$\,Pa $= 0.025 \sigma_{xx}$. Thus, lateral stresses can be neglected to a very good approximation: the elastic state of the sample is almost fully characterized by the scalar axial stress $s$.

According to Eq. \ref{se}, the volumetric and shear strains are given by $\varepsilon^{\rm pl}_{\rm V} = -\varepsilon^{\rm pl}_{\rm S} = e$, hence the plastic deformation state is completely characterized by the single scalar variable $e < 0$. The stress equilibrium condition ${\rm div}\bm\sigma =0$ implies that $\sigma = {\rm const.}(x)$. From this condition we can derive a relation between the stress $\sigma$, the plastic strain $e(x)$, and the strain $e_{\rm ext}$ that is externally imposed by the downward displacement of the top surface of the sample. Let $x$ be the material coordinate (the position in the sample prior to deformation) and let $\avop{\bullet}=L^{-1}\int\bullet\,{\rm d}x$ denote the averaging operator on $L$. We then find that 
\begin{equation}
s = E_{\rm eff} \left(e_{\rm ext} - \avop{e}\right),\quad{\rm with}\quad
E_{\rm eff}^{-1} = \avop{ E^{-1}(e(x))}
\label{Eq:s}
\end{equation}
where $E_{\rm eff}$ is the effective elastic modulus  of the sample.

The equivalent stress is $\sigma_{\rm eq} = -s$, and the yield condition reads 
\begin{equation}
f(e,s) = - s - (\Sigma_{\rm R} + \Sigma_{\rm S})\exp[- a e] - E_0 l^2 \frac{\partial^2 e}{\partial x^2} = 0\label{Eq:e}.
\end{equation}
$\Sigma_{\rm S}$ evolves according to 
\begin{equation}
\partial_t \Sigma_{\rm S} = \frac{\dot{e}}{\varepsilon_{\rm S}} \Sigma_{\rm S} + \frac{1}{\tau} (\Sigma_{\rm S}^{0} - \Sigma_{\rm S}).
\label{Eq:Sigma2}
\end{equation}
The system of integrodifferential equations for $e$, $s$ and $\Sigma_{\rm S}$ is solved as follows: In each time step we increase the externally imposed strain according to the corresponding strain rate, and evaluate $s$ and $e$ self-consistently from Eqs (\ref{Eq:s}) and (\ref{Eq:e})  with the boundary condition $\partial_x e(0) = \partial_x e(L)=0$. By comparing with the previous time step we determine the local plastic strain rates $\dot{e}$, evaluate the changes in $\Sigma_{\rm S}$ from Eq. (\ref{Eq:Sigma2}), and repeat this process until a given end strain is reached. 

\section{Model parameters}

\subsection{Elastic constants}

Material parameters of snow are often difficult to measure. This is particularly true for elastic properties, where quasi-static measurements are skewed by creep deformation, while dynamic measurements may be skewed by pore fluid effects. Furthermore, the intrinsic variability of the material leads to a huge scatter of values even if one considers a series of measurements with similar methodology and sample preparation \cite{Mellor74}. Therefore, any values of elastic constants should be considered as indicative. 

Regarding the density dependence of Young's modulus, we refer to dynamic measurements of Sigrist \cite{Sigrist06a,Sigrist06b} which cover the density range from $\rho = 200$\,kg/m$^3$ to 400\,kg/m$^3$. The data can be well represented by a power-law dependenccy, $E \propto \rho^b$ with $b = 2.94$, which compares well with the behavior of other solid foams \cite{Gibson88}. We thus adopt the value $b=3$ in our calculations. However, the absolute values obtained by Sigrist cannot be directly applied to the present context owing to the highly dynamic measurement technique. We instead refer to measurements of Schweizer and Camponovo \cite{Schweizer02} who determine shear moduli $G$ in a temperature and elastic strain rate range comparable to the present experiments. For a series of tests on natural snow with density $\rho \approx 240$\,kg/m$^3$, they give a mean shear modulus of $G = 0.75$\,MPa. With Poisson's number $\nu \approx 0$ (see below) this translates into $E \approx 2G = 1.5$\,MPa and, using the power law to extrapolate, we find for snow of density $\rho \approx 370$\,kg/m$^3$ as used in the present experiments a characteristic elastic modulus of $E \approx 5$\,MPa which we adopt as our initial value of Young's modulus. 

There are few measurements of Poisson's numbers for snow reported in the literature, with reported values being typically very small \cite{Ohizumi82}. In the present context we can obtain an upper estimate of Poisson's number from our own lateral force measurements. Identifying the side plate normal direction with the $z$ direcation and assuming that the lateral force is exclusively due to elastic cross-expansion, we find from $\sigma_{zz}/\sigma_{xx} = 0.025 = \nu/(1-\nu)$ that $\nu < 0.025$. This allows us to approximate $\nu \approx 0$ which yields the simple elastic constitutive relation $\sigma_{ij} = E \epsilon^{\rm el}_{ij}$. 

\subsection{Parameters of the plasticity model}

Fundamental to our constitutive formulation is the parameter $\alpha$ which governs the volumetric response in the plastic regime. To determine this parameter, we observe that the experimental deformation behavior is characterized by the emergence of localized deformation in the form of compaction bands that are oriented perpendicular to the compression direction. As discussed by \cite{Forest05}, such behavior occurs for $\alpha = 1/2$. We note that $\alpha = 1/2$ is also quite commonly used in deformation of metallic foams, see the parameterizations of \cite{Forest05}, \cite{Deshpande01} and \cite{Zaiser13}). Again, we can use our lateral force data to derive an upper estimate for the deviation $\delta = 1/2 - \alpha$ from this value. For $\delta \neq 0$, the lateral plastic expansion is, to lowest order in $\delta$, given by $\varepsilon_{zz}^{\rm pl}  = (2/3)\delta \varepsilon_{xx}^{\rm pl}$. Assuming that the lateral force at the end of the deformation experiment is exclusively due to the elastic accomodation of this lateral plastic expansion, and using the elastic modulus value given above to evaluate $\varepsilon_{zz}^{\rm pl} = \sigma_{zz}/E$, we find an upper estimate $\delta \le 10^{-3}$ which shows $\alpha \approx 1/2$ to be a very good approximation. 

The parameters characterizing the strain and time dependence of the yield stress in our model -- the exponent $a$ relating strength to density, the initial strength $\Sigma = \Sigma_{\rm S}^0 + \Sigma_{\rm R}$, the residual strength $\Sigma_{\rm R}$, the softening strain $\varepsilon_{\rm S}$, and the relaxation time $\tau$ -- were obtained by matching simulated and experimental stress-strain curves. This yielded the values $a = 7$, $\Sigma = 5 \times 10^4$ Pa, $\Sigma_{\rm R} = 0.25 \times 10^4$\,Pa, $\varepsilon_{\rm S} = 0.05$, and $\tau = 10$\,s. The fitted strength values are well within the range reported in the literature \cite{Mellor74,DeMontmollin82,Schweizer98}, and the same is true for the relaxation time $\tau$ \cite{Birkeland06,Szabo07}. 

\section{Stability analysis}

For $\alpha = 1/2$, instability of homogeneous uniaxial deformation with respect to formation of localized bands is associated with perturbations of strain in the axial direction, i.e., bands are oriented occur at 90 degrees to the stress axis \cite{Forest05}. Instability occurs if the derivative of the yield function with respect to the plastic strain becomes positive. Neglecting for the time being the gradient term, the stability requirement is thus:
\begin{equation}
\left.\frac{d f}{d e}\right|_t  = \frac{\partial s}{\partial e} - a (\Sigma_{\rm S} + \Sigma_{\rm R})\exp[- a e] + \frac{1}{\varepsilon_{\rm S}} \Sigma_{\rm S} \exp(-a e)   \le 0 \label{Eq:estab}.
\end{equation}
Strongly localized perturbations are of negligible influence on the stress given by Eq. \ref{Eq:s}, hence we obtain the simple stability criterion
\begin{equation}
a\varepsilon_{\rm S} \ge \frac{\Sigma_{\rm S}}{\Sigma_{\rm S} + \Sigma_{\rm R}} 
\end{equation}
This is tantamount to requiring that the rate of hardening due to densification, with structural strength fixed, exceeds the rate of softening due to bond breakage, calculated with density fixed: Densifcation-induced hardening must outweigh shear softening. 

Let us first consider the fully aged initial state. In this state, $\Sigma_{\rm S} = \Sigma_{\rm S}^0$ and (in)stability does not depend on the rate of deformation but only on the relative efficiency of hardening and softening processes in the fully aged state. For the parameters given above, we can see that the initial state of the material is within the unstable domain, hence initial deformation is in our simulations always characterized by strain localization.  

We now consider a homogeneous deformation process without spatial dependency and fixed imposed strain rate. In this case, a stationary solution of Eq. \ref{Eq:Sigma2} is $\Sigma_{\rm S} \approx \Sigma_{\rm S}^0/(1+ \dot{e}\tau/\varepsilon_{\rm S})$, i.e., strength is reduced due to the on-going destruction of bonds in the snow microstructure. This leads to an upper critical strain rate above which the specimen behaves after initial softening like a cohesionless aggregate which compacts homogeneously. From the parameters above, the critical strain rate can be estimated to be of the order of $1$ s$^{-1}$, which is well above the strain rate regime accessible in our experiments. 

\section{Correction for friction forces}

The lateral stresses measured at the end of our experiments are small, which justified the assumption of a uniaxial stress state as consistent with the parameters $\alpha = 0.5$ and $\nu \approx 0$. Nevertheless, the lateral forces cause friction forces which oppose the movement of the anvil and thus influence the measured force data. To clarify the nature of these forces we observe that, during passage of the first compaction band, there is an asymmetry of the recorded force signal depending on whether the band moves downwards or upwards. During motion of a compaction front from top to bottom of the sample, the total force rises as more snow becomes compacted and moves with the anvil. By contrast, the force remains steady when snow is being compacted but not moving with the anvil (motion of a compaction front from bottom to top). Comparing both cases allows us, together with the lateral force data, to estimate the coefficient of friction at the wall.

With $\delta=1/2-\alpha$ the lateral plastic expansion of the snow is given by $\varepsilon_{yy}^{\rm pl} = \varepsilon_{zz}^{\rm pl} = (2/3)\delta e$ where $e$ is the axial compressive strain evaluated for $\alpha = 1/2$. Because of the confinement due to the side walls, this expansion is offset by a compressive stress $\sigma_{yy} = \sigma_{zz} = (2/3)\delta e(x) E(x)$ which causes a total lateral force
\begin{equation}
\frac{2 \delta}{3} w L  \avop{e(x)E(x)} 
\end{equation}
If frictional motion occurs at the side walls, i.e., if the displacement rate $\dot{u}_x$ is negative, 
this in turn causes a total friction force 
\begin{equation}
\frac{4 \delta \mu}{3} w L \avop{\Theta(-\dot{u}_x) e(x)E(x)}
\end{equation}
where $\mu$ is the coefficient of friction at the side wall-snow interface and $\Theta$ is Heaviside's function. This force superimposes on the force required for axial compression of the snow sample and needs to be taken into account when comparing experimental and simulated force distance curves. In particular, during passage of the first band our constitutive model predicts a constant stress, hence, any slope of the measured force distance curve can be attributed to friction. From the measured slope and the lateral forces, we can deduce an estimate for the coefficient of friction at the side walls, which is found to be in the range $0.4 < \mu < 0.6$. This is in rough agreement with a direct friction measurement using a weighted block of snow sliding on the side surface of the container, which yielded $\mu \approx 0.33$.

Our simulated force distance curves shown in the main paper have been corrected for the effect of friction. With $\mu \approx 0.5$, the total friction force at the end of the experiment is of the order of 45 N which, while significantly below the total compressive force of about 250 N, is not negligible.

\bibliographystyle{plain}